# Programmable anisotropic digital metasurface for independent manipulation of dual-polarized THz waves based on voltage-controlled phase transition of VO2 microwires

Javad Shabanpour

Department of Electrical Engineering, Iran University of Science and Technology, Narmak, Tehran 16486-13114, Iran, m.javadshabanpour1372@gmail.com.

*Abstract:*

Programmable metasurfaces incorporated with tunable materials controlled by external stimuli can provide an unprecedented degree of freedom in dynamical wave manipulation in real-time. Beyond the scope of isotropic reconfigurable metasurfaces that only support unique tunable responses for excitation with a certain single-polarization, here, for the first time a new generation of ultrafast reprogrammable multi-functional anisotropic metasurface is reported to enable interchangeable missions independently for two orthogonal linearly polarized THz wavefront excitations. The reconfigurability of the proposed anisotropic meta-device was guaranteed by elaborately designed meta-particle composed of two perpendicular VO2 microwires whose operational statuses can be arbitrarily and dynamically tuned among two digital states of "0" and "1" independent for dual-polarization channels by mere changing the biasing voltage via two independent computer-programmed multichannel DC network. Capitalizing on such meta-particle design, single/multiple focused THz beam into a pre-determined focal spot and single/multiple focused vortex beams with interchangeable OAM modes satisfactorily generated by encoding a metasurface with different parabola and spiral-parabola-like coding sequences respectively. Besides controlling the near field behaviors, the versatility and flexibility of the proposed anisotropic metasurface also can furnish an inspiring platform to manipulate the far-field scattering patterns in each desired polarization channel. Consequently, two/four arbitrary oriented pencil beams, twin vortex beams with opposite OAM modes and random diffusion of terahertz wave are also realized by the proposed structure. Anisotropic meta-device bringing new degrees of freedom in achieving versatile tunable control of differently polarized electromagnetic waves which will significantly enhance storage density and data capacities and has the potential for complicated wave manipulation such as ultrafast THz communication and dynamic holography.

*1.Introduction*

Artificial metasurfaces composed of subwavelength engineered phase-shifting scatters have drawn significant attention over the years due to their capabilities to tailor the permittivity and permeability to reach values beyond material composites found in nature [1,2]. Such metasurfaces can be structured for sophisticated and practical manipulation of electromagnetic (EM) waves by purposefully controlling the local characteristics of phase, amplitude and polarization of the reflected/transmitted EM waves. The versatility and reliability of wavefront engineering by metasurfaces furnish an inspiring platform for realizing novel physics phenomena including perfect imaging [3], invisibility cloaking [4], vortex beam [5] and numerous innovative functional metadevices [6,7]. Beyond the scope of analog metasurfaces, digital metasurfaces have recently been reported as an alternative approach to engineer the scattering patterns by multiple discrete digital states ("0" or "1" state in 1-bit coding case) while they remarkably provide a broader range of wave-matter functionalities [8,9]. Despite the rapid growth, all of the aforementioned coding schemes are isotropic coding metamaterials; hence, due to their isotropic geometry of the meta-particles, these devices only support unique responses for excitation with a certain single polarization.

Since each polarization can be considered as one independent information channel in communication systems, the concept of anisotropic coding metasurface has been reported as a technique to provide new degrees of freedom to manipulate THz wavefront shaping which possesses distinct functionalities via different polarizations [10]. Despite the great progress attained by anisotropic metasurfaces [11,12], most of them cannot support reconfigurable responses and their functionalities remain fixed once they are constructed which significantly hinders their further usage in real-world practical applications. More recently, active anisotropic metasurface loaded with voltage-controlled varactors has been reported [13] whose functionalities can be dynamically tuned at microwave frequencies with real-time reconfigurability independent for two linear-polarization channels. However, as the frequency extends toward the terahertz frequency region, the active elements such as varactor/pin diodes are not commercially accessible and scaling these devices to higher frequency spectra is impossible. To solve this problem, for the first time, we proposed a multifunctional re-programmable anisotropic digital metasurface for independent dual-polarized THz wavefront engineering incorporating elaborately designed meta-particles with phase-change materials (PCM). To realize such a strategy, we have benefited from Vanadium dioxide (VO2) exotic features.

VO2 is a smart material, which is praised for its ultrafast reversible first-order phase transition from insulating monoclinic to the metallic tetragonal phase at a critical temperature Tc = 68°C [14]. This metal-insulator transition (MIT) due to atomic level deformation in VO2 can be thermally [15], optically [16] or electrically (charge injection or Joule heating) [17-19] tuned. The mechanism of MIT in VO2 is still under dynamic investigations and remains controversial as it may originate from Joule heating [20] or electric field effects [21]. Studies that were explored in [22] and [23] through time-resolved x-ray diffraction and time-resolved optical transmission disclosed that only a short time (<500 fs) is required for occurring the MIT in VO2. During the transition, the electrical resistivity has dramatic changes and can vary up to four orders of magnitude across the two phases [24]. Due to an ultrafast switching time, near room critical temperature and beneficial structural transition characteristics, VO2 has identified as considerable material in reconfigurable metamaterial devices over a broad spectral range from GHz [25] to optics [26] and has numerous functional applications at terahertz frequency region suchlike reconfigurable digital metasurface [27], polarization converter [28], reflection/transmission THz waves modulator [29] and tunable antennas [30].

Nowadays, with the rapid development of modern high data rate THz wireless communication, current applications in terahertz region are in dire need of high speed/quality communication and large data storage capacity [31]. Undoubtedly, the re-programmable anisotropic metasurfaces capable of dynamically operating distinct functionalities for excitation of different polarizations will further significantly enhance their information capabilities that are remarkably desirable from application aspects which remain largely unexplored. To fill this gap, here for the first time a reconfigurable polarization-switchable multitasking digital anisotropic metasurface with ultrafast switching between several missions at THz frequencies has been demonstrated. Each constitutive meta-atom is composed of two perpendicular VO2 microwires whose reflection phase responses can be arbitrarily and dynamically engineered in a real-time manner at will independently for two orthogonal linearly polarized (LP) excitations. Capitalizing on such elaborately designed meta-particle enabled us to satisfactorily generate single/multiple focused THz wavefront into a pre-determined point and single/multiple focused vortex beams with interchangeable OAM modes by encoding VO2 based anisotropic metasurface (VBAM) with different parabola and spiral-parabola-like coding sequences respectively. Furthermore, from the far-field perspective, any of the dual functions supported by our 1-bit VBAM structure (such as two/four arbitrary oriented pencil beams, two titled vortex beams with opposite OAM modes and random diffusion of terahertz wave) are realized by changing the polarization states of the incident THz wave between the x and y directions which is dynamically interchangeable. Several illustrative examples corroborated through numerical simulations and theoretical predictions. The proposed VBAM structure bringing a new degrees of freedom in realizing versatile reconfigurable functionalities and we believe that the straightforward and integrated structural design, ultrafast switching time, real-time multi-mission reconfigurability independent for dual-polarization channels, open up a new opportunity toward high speed/data rate wireless communication, dynamical signal processing, and future practical THz devices.

*2.Materials and methods*

**Figure 1a** represents the sketch of the proposed anisotropic programmable digital metasurface, operating independently for two linear polarized (LP) waves with dynamic THz wavefront engineering. The fundamental objective of this work is to realize a versatile VBAM meta-device in which the operational status of each occupying meta-atom can be individually addressed via two independent computer-programmed multichannel DC voltages, providing flexible and real-time control over two spatial coding sequences onto the metasurface for x- and y-LP excitations. Toward this aim, the key step is to design a bias encoded meta-particle which show two different coding states of '0' and '1,' with independent properties under the normal incident of two orthogonal polarization directions. Since such a meta-particle has anisotropic EM responses for two distinct polarization states, the reflection coefficient can be expressed by a tensor $\bar{R}_{mn}$ as follows:

$$\bar{R}_{mn} = \begin{bmatrix} \hat{x}R^x_{mn} & 0 \\ 0 & \hat{y}R^y_{mn} \end{bmatrix}$$

where $R^x_{mn}$ and $R^y_{mn}$ express the reflection coefficients of a specific meta-particle $(m.n)$ on the metasurface along two orthogonal directions. Considering $R = re^{j\varphi}$, then the phase responses under the $x$ and $y$ polarizations can be denoted by $\varphi_{xx}$ and $\varphi_{yy}$ which should be dynamically switched to possess arbitrary phase responses 0° and 180° to mimic two digital states of "0" and

"1" independently for two orthogonal LP excitations for a 1-bit programmable anisotropic metasurface.

As depicted in **Figure 1(a-b)**, the meta-particle composed of five layers, which from top to bottom, are two perpendicular VO2 microwires, substrate, gold plane (top metallic layer), substrate, two pieces of separated patches (bottom metallic layer). At the top layer, two perpendicular VO2 microwires are deposited on the sapphire as a suitable substrate due to the beneficial lattice matching effect [32] with the relative permittivity of $9.4 + 0.0001i$ and the thickness of 15 μm. The complex dielectric properties of the VO2 can be defined by Bruggeman effective medium theory in the terahertz region where $\varepsilon_d$ and $\varepsilon_m$ denote the relative permittivity of the semiconductor and metallic states and $V$ characterizes the volume fraction of the metallic state [33].

$$\varepsilon_{VO2} = \frac{1}{4}\left\{\varepsilon_d(2-3V) + \varepsilon_m(3V-1) + \sqrt{[\varepsilon_d(2-3V) + \varepsilon_m(3V-1)]^2 + 8\varepsilon_d\varepsilon_m}\right\}$$

Typical VO2 thin films that are grown on c-type or r-type sapphire substrate, in the insulating state expose electrical conductivity in the range of 10~100 S/m with the relative permittivity of about 9 [34,35]. By applying external DC bias voltage directly to metallic contacts, structural transformation occurs and VO2 turns into the metallic phase wherein the conductivity increases as high as an order of $10^5\ s/m$ [36,37]. We set $\sigma_{off} = 10\ s/m$ ( "0" digital state) and $\sigma_{on} = 5 \times 10^5\ s/m$ ( "1" digital state) in the insulating and metallic regions corresponding to $T_C = 300K$ and $T_H = 400K$ respectively. Other geometrical parameters optimized as $P = 50\ \mu m$, $T = 30\ \mu m$, $h_1 = 15\ \mu m$, $w = 4\ \mu m$. The conductivity of VO2 microwire along x-direction denoted by $\sigma_x$ while that along y-direction is denoted by $\sigma_y$ which can be individually programmed by using field-programmable gate array (FPGA) hardware. Two lateral metallization electrodes for each VO2 microwires are patterned using aligned e-beam lithography followed by e-beam evaporation and lift-off process of a thin gold film to form an electrical contact. Practical realization of independently and locally programming each VO2 microwires simultaneously through FPGA network platform can be described as follows: two metallic via holes are drilled through the substrate to connect one of the two electrical contacts for each VO2 microwires (marked with a negative sign in the picture) and the second metallic layer (ground plane) as a negative electrode. Two metallic via holes are drilled through the substrates to attach other electrical contacts (marked with a positive sign in the picture) with two pieces of separated patches on the bottom metallic layer which act as positive electrodes to apply the DC bias voltage that are electrically isolated from the negative electrode.

For evaluating the reflection characteristics of the proposed meta-particle, we perform full-wave simulations with the commercial CST software when periodic boundary conditions are utilized along the x and y directions and open boundary condition is applied along z-axis which shows opposite coding states with ~ 180° phase differences in a wide frequency band (**see figure 1c**). Therefore four arbitrarily reflection phases of $\varphi_{xx}/\varphi_{yy}$ = 0°/0°, 0°/180°, 180°/0°, and 180°/180° can be independently obtained to mimic four digital states of "0/0," "0/1," "1/0," and "1/1" where binary codes before and after the slash symbol (/) represent the digital states under the illumination of x- and y-LP wave, respectively. By adjusting the electrical conductivity of the VO2 microwires in the proposed meta-particle to $\sigma_x/\sigma_y$ = 10 / 10, 10 / 5 × 10$^5$, 5 × 10$^5$/ 10, 5 × 10$^5$/ 5 × 10$^5$, we can correspondingly obtain the coding states of "0/0", "0/1", "1/0", and "1/1" which can be dynamically tuned by applying the DC bias voltage.

### 3. Result and discussion

For a quantitative illustration of VBAM structure, we first encode a metasurface with spatial phase gradients along y-axis and chessboard configuration along x-axis enabling two and four symmetrically oriented scattered beams respectively as demonstrated by the 3D scattered far-field patterns in **Figure 2a**. All of the multi-mission provided with the proposed VBAM structure also can be dynamically interchanged along both x and y polarizations by mere changing the biasing voltage as depicted in **Figure 2b**. The encoded metasurface containing $32 \times 32$ meta-particles with the overall size of $1.2 \times 1.2 \ mm^2$. The deflected angle can then be calculated from generalized Snell's law as $\sin^{-1}(\lambda/2MP)$ and $\sin^{-1}(\sqrt{2}\lambda/2MP)$ for two and four symmetrically oriented scattered beams respectively where $\lambda$ represents the working wavelength at the center frequency of 1.5 THz and $M$ demonstrates the number of meta-particles in each lattice. obviously, by changing the dimension of the lattices one can adjust the deviation angle in a real time manner independent for orthogonal polarizations and beam steering functionality could be envisioned In all cases, the simulated deviation angles have good conformity with our theoretical predictions. As a new scenario, the coding sequences can be arranged in such a way to deflect the y-polarized terahertz beam to two symmetrical directions using coding sequence of (0101…/0101…).

Simultaneously, when coding particles are randomly distributed on the surface along x direction, diffusive scattering can be invoked to randomly redirect the x-polarized beam into numerous directions using an optimized '0' and '1' coding sequence as depicted in **Figure 3(a-b).** These examples demonstrate the capability and flexibility of the proposed multifunctional VBAM to independently manipulate the x- and y-polarized incident terahertz waves in a real -time manner.

The versatility and flexibility of the proposed anisotropic metasurface also can furnish an inspiring platform to utilize it as a metalens under each desired orthogonal polarizations to concentrate the incident THz wavefront into a predetermined focal spot which can be dynamically altered. To engineer this feature of the work, the required parabola phase distribution can be calculated as [38]:

$$\varphi(x.y) = \frac{\omega}{c}\left(\sqrt{r^2 + z_f^2} - z_f\right)$$

where $\omega$, $r$, and $z_f$ are the angular frequency, radial coordinate, and focal length respectively which in the cartesian coordinate $r^2 = (x - x_f)^2 + (y - y_f)^2$. Let us consider VBAM driven by parabola phase distribution in such a way to concentrate incident THz wave at $x = 200$ $\mu m$, $y = -200$ $\mu m$ and $z = 270$ $\mu m$ for illuminating y-polarized incident plane wave whiles capable of focusing the reflected energy under x-polarized incidence at $x = -100$ $\mu m$, $y = 300$ $\mu m$ and $z = 270$ $\mu m$. Simulated normalized electric field distribution in the sampling planes at the corresponding focal length have been demonstrated in **Figure 4(a-b)**. As can be deduced from **Figure 4**, the electric field concentrations corroborate well with the predetermined focal points and some slight side lobes around the focal spot can be attributed to the finite and discrete nature of the proposed anisotropic metasurface.

To further clarify the powerful capability of controlling the near field behaviors in one polarization channel and manipulating the far-field scattering patterns in the other polarization channel simultaneously and with real-time reconfigurability, we encode the VBAM to randomly redirect the x-polarized THz wave to all direction while at the same time, concentrate the reflected energy into two focal points for y-polarized excitation. The 3D scattered far-field pattern and normalized electric near-field distribution, as well as the spatial phase profile on the metasurface, are depicted in **Figure 5(a-b)**. To engineer multiple focus feature of the work, we have divided the metasurface to multi segments in which each segment generates one focal point in space. Hence, we have driven VBAM by the required phase profile which calculated to concentrate the reflected energy at $x = -300\ \mu m$, $y = -400\ \mu m$ and $z = 100\ \mu m$ for the first segment and $x = 200\ \mu m$, $y = 400\ \mu m$ and $z = 100\ \mu m$ for the second segment. The whole space scanning characteristic of the proposed anisotropic metasurface was successfully clarified with this illustrative example which has promising perspectives in applications of reconfigurable compact metalens for imaging [39] and near-field communication [40].

Consequently, the above numerical results demonstrate that the proposed reconfigurable VBAM meta-device can successfully perform the multiple missions that were assigned to it independently for two orthogonal polarization channels. Once again, we should notice that all the practical functionalities discussed throughout this paper under one polarization channel, could be instantaneously interchanged to the other polarization state by digitally and dynamically change the spatial voltage distribution provided by two independent computer-programmed multichannel DC voltages.

## *4.Conclusion:*

Beyond the scope of fixed-function and isotropic coding metasurfaces, with the aim of integrating multiple diversified functionalities into one single device with tunable responses for two orthogonal polarization channels, we proposed anisotropic digital metasurface whose reflection behaviors can be electrically and independently tuned for two orthogonal polarized THz waves. We have benefited from VO2 fast and fruitful semiconductor to metal phase transition to design the proposed meta-particle in such a way to write two completely different coding sequences on a single metasurface which can be dynamically switched between several missions in real-time. By encoding the VBAM with chessboard phase distribution, beam splitter scattering pattern was successfully generated which is of paramount importance in the terahertz regime which can be utilized to demultiplex the THz signal modulated with two orthogonal polarization channels leads to improving the transmission rate in which data capacity can be doubled. The versatility of the proposed VBAM structure to control the near-field behavior was clarified with different illustrative examples from focused normal beam to focused vortex beam into a predetermined focal point which can be dynamically altered by tuning the DC bias voltage controlled by an FPGA platform. Moreover, from the far-field perspective, two/four arbitrary oriented pencil beams, twin vortex beams with opposite OAM modes and random diffusion of terahertz wave are also realized by the proposed structure. The additional degrees of freedom of the VBAM structure will afford powerful capabilities in manipulating both near-field and far-field behaviors simultaneously into each desired orthogonal polarization channels with interchangeable missions in real-time. With independent dual-polarization THz wavefront engineering with real-time reconfigurability, we believe that the proposed ultrafast reconfigurable THz meta-device meets well the THz future smart, intelligent, and industrial demands.

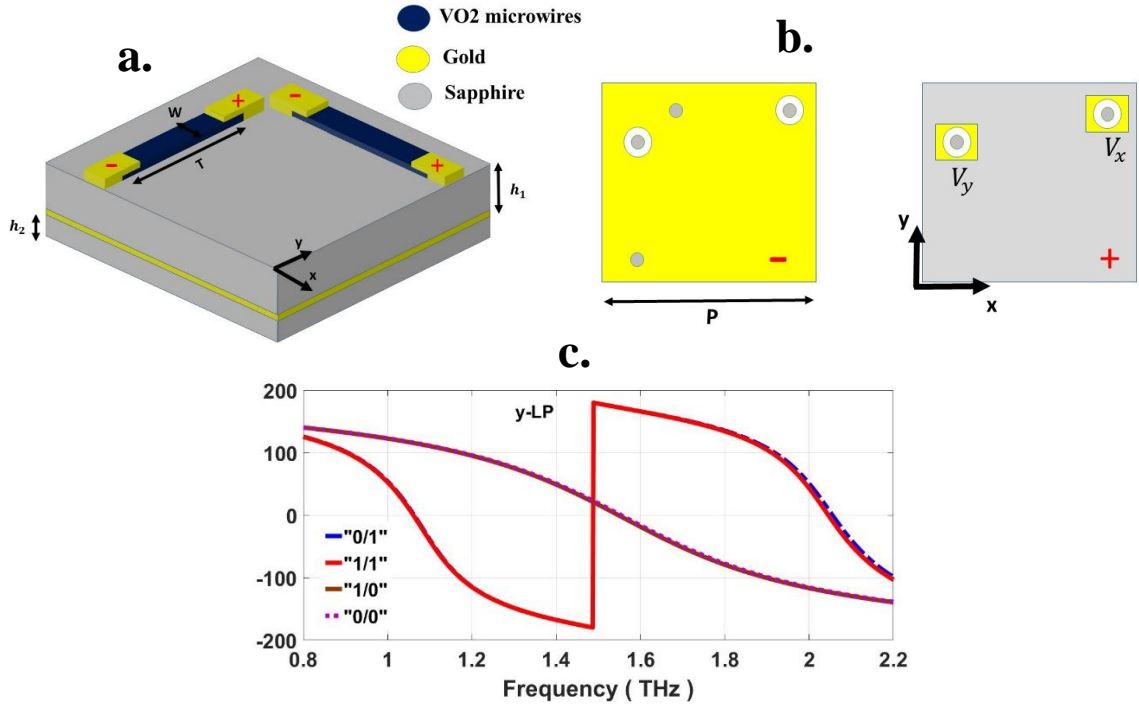

Figure 1. (a) Schematic of the active anisotropic metasurface meta-particle composing VO2 microwires with independent properties under the normal incident of two orthogonal polarization directions (b) first and second metallic layer used as negative and positive electrodes respectively (c) Reflection phases of the 1-bit anisotropic digital elements under illuminating y-polarized normal incident plane wave.

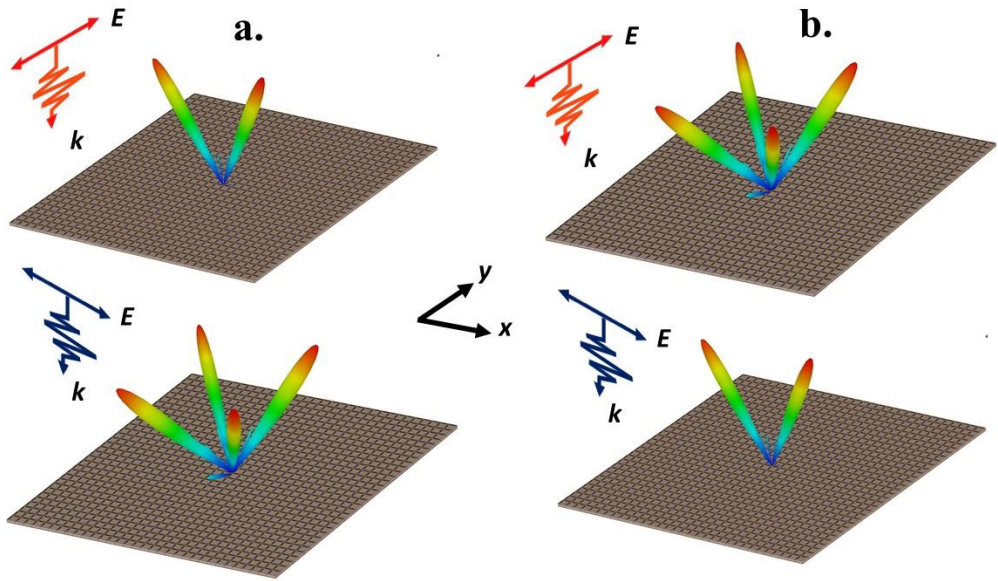

Figure 2. 3D scattered far-field patterns with spatial phase gradients along y-axis and chessboard configuration along x-axis enabling two and four symmetrically oriented scattered beams and vice versa.

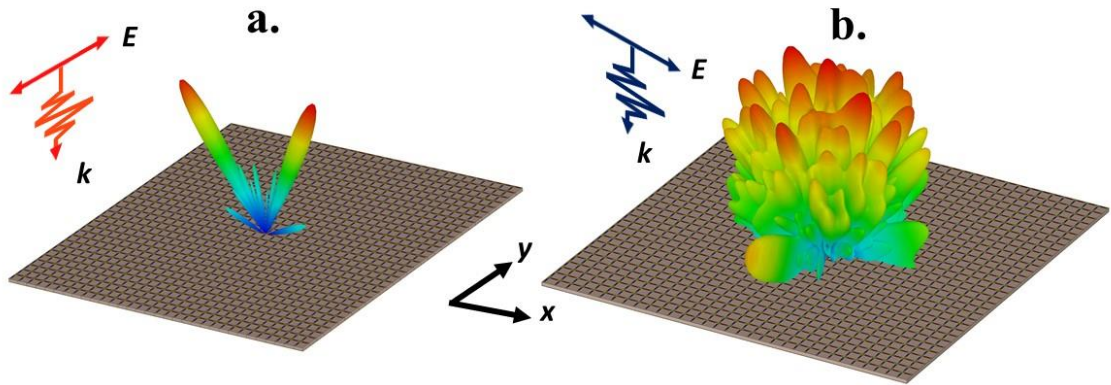

Figure 3. (a) 3D scattered far-field patterns with spatial phase gradients along y-axis to deflect the y-polarized terahertz beam to two symmetrical directions using coding sequence of (0101…/0101…). (b) when coding particles are randomly distributed on the surface along x direction, diffusive scattering can be invoked to randomly redirect the x-polarized beam into numerous directions using an optimized '0' and '1' coding sequence

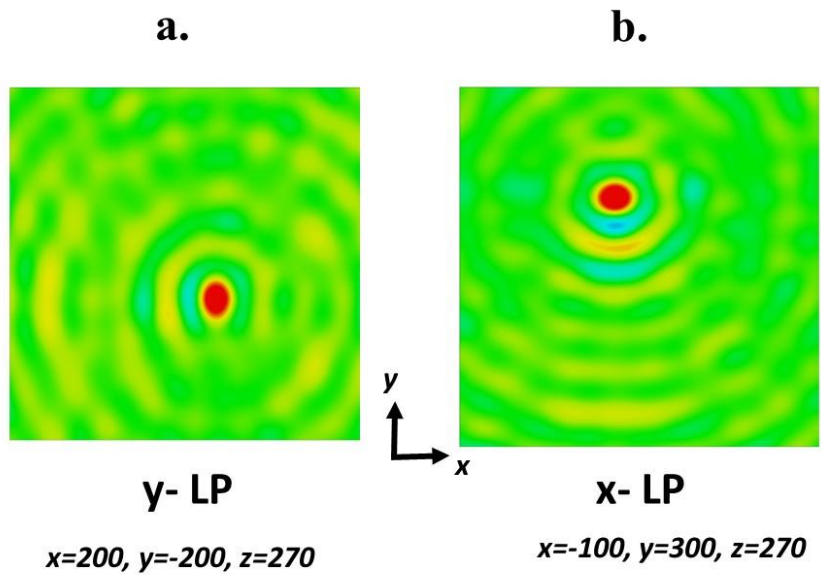

Figure 4. (a) VBAM driven by parabola phase distribution in such a way to concentrate incident THz wave at x = 200 μm, y = -200 μm and z = 270 μm for illuminating y-polarized incident plane wave and (b) focusing the reflected energy under x-polarized incidence at x = -100 μm, y = 300 μm and z = 270 μm.

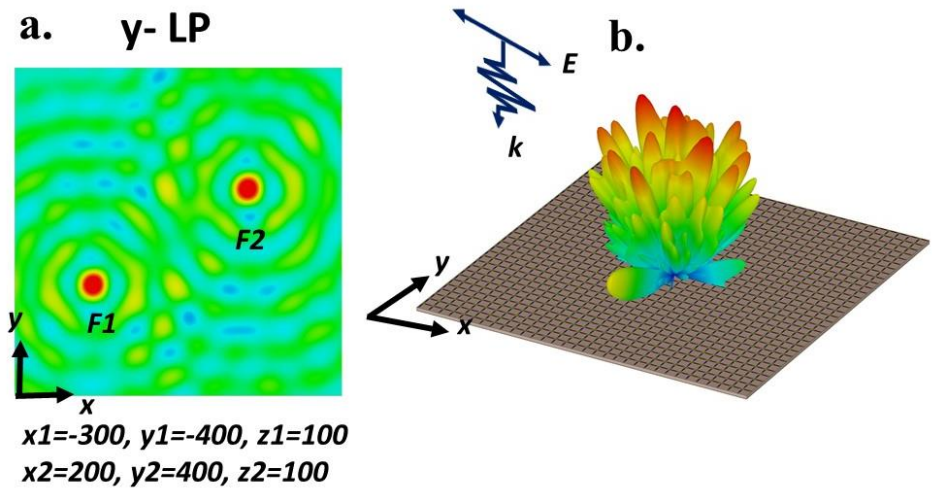

Figure 5. (a) concentrate the reflected energy into two focal points for y-polarized excitation (b) when coding particles are randomly distributed on the surface along x direction, diffusive scattering can be invoked to randomly redirect the x-polarized beam into numerous directions using an optimized '0' and '1' coding sequence.